\documentstyle[epsbox,PASJadd]{PASJ95}
\newcommand{\vol}{51}
\newcommand{\no}{4}
\newcommand{\lett}{}
\newcommand{\spage}{}
\newcommand{\rdate}{1999 March 24}
\newcommand{\adate}{1999 May 5}

\begin{document}

\title{Core--Halo Structure of a Chemically Homogeneous Massive Star and 
        Bending of the Zero--Age Main--Sequence}

\author{Mie {\sc Ishii},$^1$  Munetaka {\sc Ueno},$^{2}$ 
and Mariko {\sc Kato}$^1$ 
\\[12pt]
$^1$ {\it Department of Astronomy, Keio University, Hiyoshi,
Kouhoku-ku, Yokohama 223-8521}
\\ {\it E-mail(MK): mariko@educ.cc.keio.ac.jp }\\
$^{2}$ {\it Department of Earth Science and Astronomy, 
College of Arts and Sciences, The University of Tokyo}\\
{\it 3-8-1 Komaba, Meguro-ku, Tokyo 153-8902}
}


\abst{
We have recalculated the interior structure of
very massive stars of uniform chemical composition with the OPAL opacity.
Very massive stars are found to develop a core--halo structure 
with an extended radiative--envelope. With the core--halo structure, 
because a more massive star has a more extended envelope, 
the track of the upper zero--age main-sequence
(ZAMS) curves redward in the H--R diagram at $> 100 M_{\odot}$
($Z=0.02$), $>70 M_{\odot}$ ($Z=0.05$), and
$> 15 M_{\odot}$ for helium ZAMS ($X=0.$, $Z$=0.02).
Therefore, the effective temperatures of very massive ZAMS stars are
rather low: e.g., for a $200 M_{\odot}$ star,  
$\log T_{\rm eff}=4.75$ ($Z=0.004$),
4.60 ($Z=0.02$), 4.46 ($Z=0.05$), and 4.32 ($Z=0.10$). 
The effective temperatures of very luminous stars ($> 120 M_{\odot}$) 
found in the LMC, the SMC, and the Galaxy are discussed in 
relation to this metal dependence of a curving upper main-sequence.  
}

\kword{Stars: abundances --- Stars: individual(the Pistol star)
--- Stars: interiors --- Stars: massive --- Stars: supergiants
}

\maketitle
\thispagestyle{headings}

\section
{Introduction}

Recent infrared observations have found a number of very luminous stars in
young clusters near to the galactic center, 
in 30 Doradus in the LMC and in star-forming regions in the SMC 
(Nagata et al. 1993, 1995; 
Najarro et al. 1997; Messey, Hunter 1998). 
The Pistol star, a member of the AFGL 2004 young cluster 
(Nagata et al. 1993; Figer et al. 1996), is
one of the brightest stars known in the local group of galaxies.  
The luminosity and the temperature of the Pistol star were derived 
to be $\log L(L_{\odot})=6.6$ and $\log T_{\rm eff}(K)= 4.15$, 
respectively, from a near--infrared data analysis (Figer et al. 1998). 
Very luminous stars with such a low surface temperature are 
rarely found in conventional observations, 
which stimulates a theoretical interest in its
evolutionary stage. Compared with the evolutionary path 
of very massive stars in the H--R diagram,
Figer et al. have derived the initial mass of the Pistol star to be 
200--250$ M_{\odot}$ and the age to be 1.7--2.1 Myr. 
In the LMC, there also found a number of very luminous O3 stars
in the R136 cluster in 30 Dor, several of which 
have a mass in excess of 120 $M_{\odot}$ (Messey, Hunter 1998).
One of them,  Melnick 42, is analyzed in detail by Pauldrach et al. 
(1994), who estimate the luminosity and the mass of the star to be 
$\log L(L_{\odot})=6.6$ and $M=150 M_{\odot}$. 

In this way, very luminous stars are found in various
chemical circumstances. Their inferred mass, age, and evolutionary status
provide fundamental information for studies of star--forming regions.  
In order to estimate their age and mass, we require theoretical tracks 
of very massive ( $>120 M_{\odot}$) MS stars of both Populations I and II. 
With the discovery of the very massive stars  ($L >10^6 L_{\odot}$) mentioned
above, we need to reanalyze the structure of very massive MS stars.

Kato first found from an analysis that very massive stars develop 
a core--halo structure, which 
results in a redward bending of the upper main--sequence. 
Kato (1985) calculated the structures of very 
massive Newtonian stars   
with Compton--scattering opacity, and found that vary massive stars develop 
an extremely extended radiative envelope. Such a structure is caused by 
an outward increase of the opacity in the
radiative region.  With the Compton--scattering opacity, 
this core--halo structure appears only in very massive stars 
($> 10^6 M_{\odot}$). 
When a star develops a core--halo structure, the photospheric temperature
decreases owing to the extended envelope. 
Because a more massive star has a more extended
envelope, the upper part of the main--sequence 
curves to the right in the H--R diagram at very high luminosity, 
$L > 10^{10} L_{\odot}$, for Pop I (Kato 1986). Such a massive 
star of $\sim 10^6 M_{\odot}$ may be  a theoretical problem rather 
than a realistic one; but after  
the new opacity, we expect that the core--halo structure and 
the resultant main--sequence bending will be realized 
in much less--massive stars. This is because the Compton--scattering 
opacity increases outward only a few percent in the 
radiative region, whereas the new
opacity has a prominent peak at around $T \sim 2 \times 10^5$ K 
that must cause the structure to change effectively. 

The bending of the main--sequence appears for a much smaller mass in
the case of helium stars with the Los Alamos radiative opacity.
Langer (1989) has calculated chemically uniform helium stars with
central helium burning as models of Wolf--Rayet stars.
The pure helium main--sequence ($Y=1.0$) only indicates 
bending at $> 60 M_{\odot}$, but sequences of C/O--rich helium stars
curve rightward in less--massive stars (e.g., at $\geq 15 M_{\odot}$ for
$Y=0.02, C+O=0.98$). The bending appears in less--massive C/O--rich
stars because of a larger radiative opacity of C/O--rich  matter.

The core--halo structure is also reported in helium main--sequence
stars with optically thick winds.  Kato and Iben (1992) have presented 
the interior structure of helium main--sequence stars 
with artificially enhanced opacity in a model of 
Wolf--Rayet stars. The core--halo structure is developed in 
stars of mass 15 -- 30 $M_{\odot}$, 
in which the radiative envelope is extended in a way that the density 
profile changes while responding to any opacity variation.
In the H--R diagram, the main--sequence runs from the 
lower--left to the upper--right, 
contrary to the usual main--sequence. 
This main--sequence corresponds to the upper half part of the 
curved main--sequence described above. 

In this way, chemically homogeneous stars show a core--halo structure 
which results in the curved main--sequence 
when the opacity monotonically increases outward.
The new opacity, which varies around a prominent peak, 
requires a recalculation of 
the core--halo structure and bending of the main--sequences. 
Therefore, 
we have recalculated the interior structure of 
very massive main--sequence stars with the OPAL opacity. 
After the new opacity appeared, there have been presented many 
calculations on massive--star evolution for various problems,  
such as evolution with wind mass--loss, instability against radial
oscillation, WR star modeling, 
evolutionary connection between O stars, LBV, and WR stars etc. 
(e.g., Schaller et al. 1992; Heger, Langer 1996; Glatzel, Kiriakidis 1998;
Stothers, Chin 1996, 1997, and references therein);
but little attention has been paid to the core--halo structure
and the resultant bending of the main--sequence. 
The main--sequence up to 120 $M_{\odot}$ shows no indication 
to turn to the right for $Z \leq 0.02$, (Schaller et al. 1992), 
and it slightly does so for $Z=0.03$ (Glatzel, Kiriakidis 1993); 
however, more massive stars up to $ 300 M_{\odot}$, ZAMS curves 
redward at $ \geq 120 M_{\odot}$ for $Z=0.02$ and 0.04 
(Figer et al. 1998). In the present paper,  
we have concentrated on massive 
zero--age main--sequence stars 
in order to examine the basic properties of the core--halo structure 
and to confirm the bending of MS. The next section describes the 
method and assumptions of the numerical calculation, and section 3 
shows the core--halo structure of chemically homogeneous stars and 
presents curved ZAMS for Populations I and II
in the H--R diagram. A comparison with observational data of very 
luminous stars, such as the Pistol star, is given in discussions.

\section
{Calculations}

The structure of massive stars with uniform chemical composition 
is obtained by solving the equations of hydrostatic balance, mass 
continuity, energy transfer by diffusion and by convection, and energy 
conservation with the assumption of spherical symmetry. 
We calculated two sets of ZAMS solutions, i.e., with/without mass--loss. 
The wind mass--loss is assumed to be in the quasi--steady state  
in which the heat flux is steady in the $q$--coordinate,  
i.e., the gravitational contraction term in the energy--conservation 
equation is approximated by 
$ \varepsilon_{\rm g}=-T(\dot M /M)(\delta s/\delta \ln~q)_{t}$,  
where $q=M_r /M$ is the mass within radius $r$ divided
by the stellar mass, and the suffix $t$ denotes the partial derivative 
with constant time. These equations and assumptions are essentially 
the same as those in Kato (1980).
This quasi--steady state condition stands well for a chemically 
uniform star with mild mass--loss. We have checked it by reproducing 
the diffusive luminosity distribution obtained by a hydrodynamic 
calculation for a 15 $M_{\odot}$ He MS star (Heger, Langer 1996).
The wind mass--loss rate is assumed to be zero and $5 \times 10^{-5} 
M_{\odot}$yr$^{-1}$ for MS stars.
Because of the effects of wind mass--loss on the stellar structure 
is very small, the interior structure hardly changes up to 
$1 \times 10^{-4} M_{\odot}$yr$^{-1}$: the effective temperature 
increases only by $ \Delta \log T_{\rm eff} = 0.003$  
if we include a mass loss of $\dot M = 1 \times 10^{-4} M_{\odot}$yr 
$^{-1}$ in the 200 
$M_{\odot}$ model.
The updated OPAL opacity (Iglesias, Rogers 1996) is used and 
the mixing--length parameter for convective energy transport is 
set to be 1.5.
The chemical composition of stars is assumed to be uniform 
with $X=0.7$ for hydrogen, and $Z=0.004, 0.02, 0.05,$ and 0.1 for heavy 
elements by weight. In addition to these compositions, 
we have calculated additional models with different 
combination for a comparison: $X=0.35$ and $Z=0.05$, and helium ZAMS of 
$X=0$, and $Z=0.004, 0.02, 0.05,$ and 0.1 without mass loss. 

\begin{figure}[t]
\begin{center}
\epsfile{file=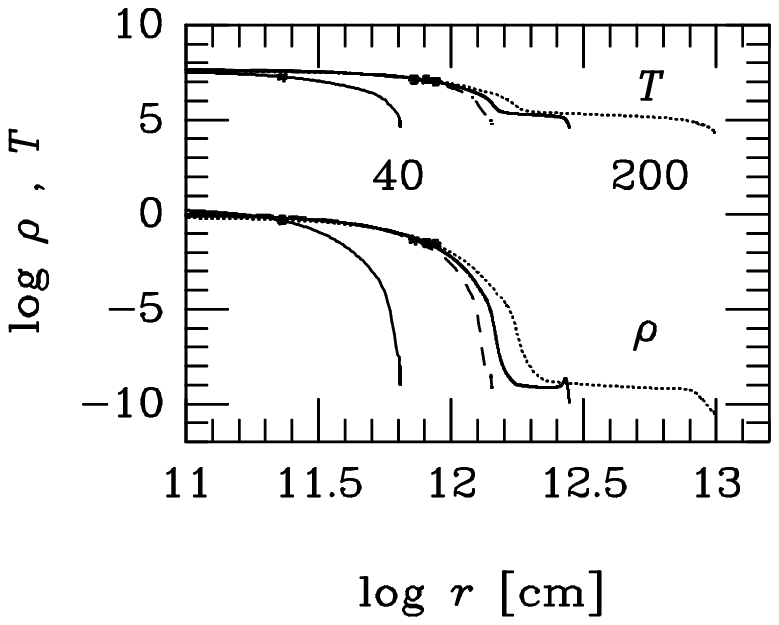,width=7cm,height=5cm }
\end{center}
\caption
{Temperature and density distributions of chemically uniform stars
($X=0.7, Z=0.02$) of 40 and 200  $M_{\odot}$ (solid curve).
The dot denotes the outer edge of the convective core. The outermost
point of each curve denotes the photosphere. The dashed and dotted curves
denote the stars of $200 M_\odot$ with ($X=0.7,
Z=0.004$) and ($X=0.7,Z=0.10$), respectively.
}
\end{figure}

\section
{Interior Structure of Massive Stars and the H--R Diagram}


Figure 1 shows the distributions of the density and the temperature 
of chemically homogeneous stars of  
40 and $200 M_\odot$ for $Z=0.02$. 
The $200 M_\odot$ star has a quite different structure from the
$40 M_\odot$,  because it develops an extended isothermal radiative--envelope 
where the density is almost constant. Such a core--halo structure  
develops well in very massive stars ($> 150 M_\odot$), but does not do so 
in less--massive stars, such as the $40 M_\odot$, as shown in this figure. 
These two different types of homogeneous stars have already been 
pointed out by Kato (1985). 
Following her way, we call the core--halo structure a Type--II solution,  
and for the other one, the usual structure like in a $40 M_\odot$ star, 
a Type--I solution. 

\begin{figure}
\begin{center}
\epsfile{file=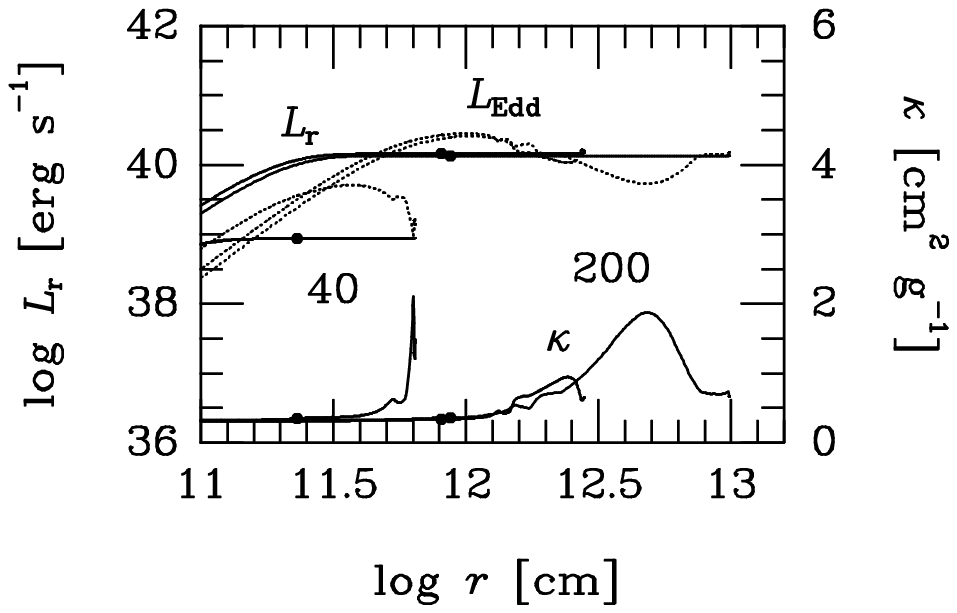,width=7cm,height=5cm }
\end{center}
\caption
{Variation in the diffusive luminosity $L_{r}$,
the local Eddington luminosity,  $L_{\rm Edd}= 4\pi c G M_{r} /\kappa$,
and the opacity $\kappa$ for models in figure 1, except the
  $200 M_\odot$ with ($X=0.7,Z=0.004$), which is omitted from this figure.
 $L_{r}$, solid curve; $L_{\rm Edd}$, dotted curve; $\kappa$, lower solid
curve.
}
\end{figure}

Figure 2 shows the distributions of the diffusive luminosity, the local
Eddington luminosity, 

\begin{equation}
\label{eq:edd}
$$L_{\rm Edd} = {{4 \pi cG M_r}\over{\kappa}},$$
\end{equation}

\noindent
and the opacity. 
As shown in this figure, the local Eddington luminosity decreases outward
in the radiative region,  
corresponding to an increase of the OPAL opacity toward the peak at $T 
\sim 2 \times 10^5$ K. 
The $200 M_\odot$ star shows the super--Eddington luminosity in the outer
radiative region, where the convective heat transport is inefficient and 
a wide isothermal region develops to form a core--halo structure,  
as shown in figure 1.  In the $40 M_\odot$ model, 
the diffusive luminosity does not reach the Eddington   
luminosity in the radiative region, and thus no core--halo 
structure appears. 

We also calculated the structures of helium ZAMS stars ($X=0.0$) 
and helium--rich stars ($X=0.35$, $Z=0.05$). Type--II solutions also appear 
in these stars, and their basic properties are the 
same as those in ZAMS stars. 
Figure 3 shows the density and the temperature distributions
for helium stars of 12 and $40 M_\odot$; 
the latter shows the core--halo structure. 

\begin{figure}
\begin{center}
\epsfile{file=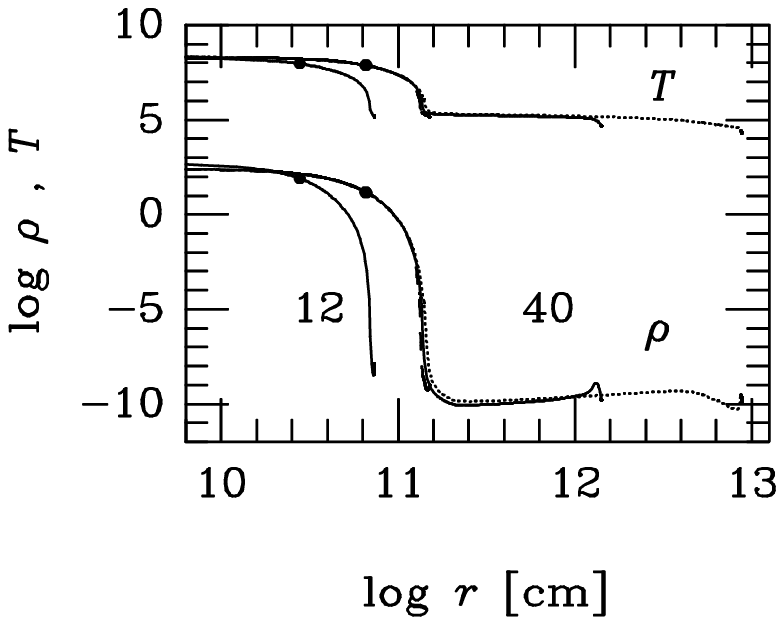,width=7cm,height=5cm }
\end{center}
\caption
{
Same as those in figure 1, but for He main--sequence stars of
12 and $40 M_\odot$ with ($X=0., Z=0.02$). The thin dashed and dotted curves
denote the stars of $40 M_\odot$ with
$Z=0.004$ and $Z=0.05$, respectively.
}
\end{figure}

Such a core--halo structure is more extended in metal--rich stars. 
Figure 1 demonstrates the difference of interior structures 
of a $200 M_\odot$ star with various metal contents, 
$Z=0.004, 0.02 $, and $0.10$.
Metal--rich stars develop a wide isothermal 
radiative envelope. 
As the envelope is most extended, the effective temperature of the star
is the lowest of these three stars.

\begin{table*}
\small
\begin{center}
Table~1.\hspace{4pt}Characteristic values of very massive star.\\
\end{center}
\vspace{6pt}
\begin{tabular*}{\textwidth}{@{\hspace{\tabcolsep}
\extracolsep{\fill}}p{5pc}clccclcc}
\hline\hline\\[-6pt]
Mass $(M_{\odot})$ & $X$ & $Z$ & $ R_{\rm ph}(R_{\odot})$ & $\log
T_{\rm ph}$(K) & $\log L_{\rm ph}/L_{\odot}$
 &  $L_{\rm ph}/L_{\rm Edd}^*$  &  1-$(M_{\rm c}/ M )^\dagger$ & $\log T_{\rm c}$ (K)\\   
[4pt]\hline\\[-6pt]
40 \dotfill   & 0.7 & 0.02 & 9.3 & 4.62 & 5.35  & 0.65 & 0.38 & 7.57\\
60 \dotfill   & 0.7 & 0.02 & 12  & 4.65 & 5.71  & 0.74 & 0.28 & 7.59 \\
100 \dotfill  & 0.7 & 0.02 & 18 & 4.67 & 6.11  & 0.84 & 0.19 & 7.61 \\
200 \dotfill  & 0.7 & 0.02 & 40 & 4.60 & 6.57  & 0.92 &0.091 &7.63  \\
250 \dotfill  & 0.7 & 0.02 & 55 & 4.57 & 6.71  & 0.95 &0.077 &7.63\\
300 \dotfill  & 0.7 & 0.02 & 71 & 4.54 & 6.82  & 0.96 &0.066 &7.63 \\
500 \dotfill  & 0.7 & 0.02 & 160 & 4.44 & 7.10  & 0.97 &0.042 & 7.65\\
1000 \dotfill & 0.7 & 0.02 & 3000 & 4.08 & 7.46  & 0.95 &0.029 &7.66 \\
100 \dotfill & 0.7  & 0.05 & 26  & 4.58  & 6.09  & 0.87  & 0.19 & 7.58 \\
200 \dotfill  & 0.7 & 0.004 & 20  & 4.75 & 6.58  & 0.87 &0.096 &7.66 \\
200 \dotfill  &  0.7 & 0.05 & 76 & 4.46 &6.56 &0.93  & 0.091  &7.60 \\
200 \dotfill  & 0.7 & 0.1 & 140 & 4.32  &6.55  &0.89  &0.10  &7.58 \\
40 \dotfill & 0.35 & 0.05 &  15 & 4.59 & 5.68 & 0.85 & 0.24 & 7.58\\
100 \dotfill & 0.35 & 0.05 & 75 & 4.41 & 6.34 & 0.86 & 0.10 & 7.62\\
150 \dotfill & 0.35 & 0.05 & 200  & 4.26 & 6.59 & 0.89 & 0.073 & 7.63\\
8  \dotfill  &  0. & 0.02 & 0.82 & 5.04 & 4.94 & 0.73 & 0.41& 8.13 \\
12  \dotfill & 0. & 0.02 & 1.1 & 5.07 & 5.29 & 0.88 &0.32 &8.28 \\
15  \dotfill  &  0. & 0.02 & 1.3 & 5.08 & 5.47 & 0.93 &0.28 &8.30 \\
30  \dotfill  &  0. & 0.02 & 8.3 & 4.79 & 5.96 & 0.94 & 0.18&8.31 \\
40 \dotfill & 0. & 0.02 &   20  & 4.64 & 6.14 & 0.97 & 0.15 & 8.32\\
60  \dotfill  &  0. & 0.02 & 71 & 4.43 & 6.38 & 0.80 &0.12 &8.33 \\
100 \dotfill  &  0. & 0.02 & 670 & 4.02 & 6.67 & 0.02 & 0.096 &8.36 \\
130 \dotfill  &  0. & 0.02 & 850 & 4.00 & 6.81 & 0.017&0.087 &8.36 \\
40 \dotfill & 0. & 0.004 &  2.2 & 5.13 & 6.14 & 0.99 & 0.15 & 8.32\\
40 \dotfill & 0. & 0.05 &  130  & 4.25 & 6.14 & 0.81 & 0.15 & 8.32\\
[4pt]
\hline
\end{tabular*}

\vspace{6pt}
\noindent
$^*$ The ratio of the photospheric luminosity to the Eddington luminosity at
the surface.\\
\noindent
$^\dagger$ The ratio of the mass of the radiative envelope to the total stellar mass.
\end{table*}

Table 1 gives the physical quantities of solutions for very
massive stars with various sets of chemical composition, 
$X$ and $Z$. The four columns next to $Z$ give the photospheric values of the 
radius $R_{\rm ph}$, temperature $T_{\rm ph}$, luminosity 
$L_{\rm ph}$, and ratio of the diffusive 
luminosity to the Eddington luminosity at the photosphere  
$L_{\rm ph}/L_{\rm Edd}$. 
The next gives the ratio of the mass of the radiative
envelope to the total stellar mass, $1-(M_{\rm c}/M)$, where $M_{\rm c}$
denotes the mass of the convective core. The last column gives 
the central temperature $T_{\rm c}$ of the star.

In the main--sequence with ($X=0.7$, $Z=0.02$) in table 1, 
$T_{\rm ph}$ increases with the stellar mass until it has the maximum value 
at $\sim 100 M_\odot$, and decreases after that as the core--halo structure 
develops.  Here, we define the critical mass which divides the Type--I 
and Type--II solutions as the stellar mass of the maximum 
effective temperature in each sequence. 
This critical mass is tabulated in tables 2 and 3, for ZAMS and He--ZAMS, 
respectively. 

\begin{table}
\small
\begin{center}
Table~2.\hspace{4pt}Critical mass between Type I and Type
II solutions of ZAMS star.\\
\end{center}
\vspace{6pt}
\begin{tabular*}{\columnwidth}{@{\hspace{\tabcolsep}
\extracolsep{\fill}}p{6pc}ccc}
\hline\hline\\[-6pt]
$Z$ &  Mass ($M_{\odot}$)  & $\log T_{\rm eff}(K)$ & $\log L_{\rm ph}
(L_{\odot}$) \\
[4pt]\hline\\[-6pt]
      0.004   \dotfill    &     200    &     4.75   &  6.58\\
      0.02     \dotfill &       90    &     4.67   &  6.03\\
      0.05     \dotfill &       70    &     4.60   & 5.81\\
      0.1      \dotfill  &       60     &     4.54   & 5.66  \\[4pt]
\hline
\end{tabular*}
\vspace{6pt}

\noindent
\end{table} 

\begin{table}
\small
\begin{center}
Table~3.\hspace{4pt}Critical mass between Type I and Type II solutions 
of Helium ZAMS stars.\\
\end{center}
\vspace{6pt}
\begin{tabular*}{\columnwidth}{@{\hspace{\tabcolsep}
\extracolsep{\fill}}p{6pc}ccc}
\hline\hline\\[-6pt]
$Z$ &       Mass ($M_{\odot}$)  & $\log T_{\rm eff}$(K) &
 $\log L_{\rm ph}(L_{\odot}$) \\
[4pt]\hline\\[-6pt]
      0.004  \dotfill    &   30     &  5.13    & 5.96  \\
      0.02   \dotfill   &       15    &     5.07   & 5.29\\
      0.05   \dotfill   &       10    &     5.03   & 5.14\\   
      0.1    \dotfill    &      7     &     4.98   & 4.83  \\[4pt]
\hline
\end{tabular*}
\vspace{6pt}

\noindent  

\end{table}


Figure 4 shows four tracks of ZAMS with $X=0.7$, and
$Z=0.004,0.02,0.05$, and 0.10, left to right, in the H--R diagram. 
The metal--poor main--sequence ($Z=0.004$) curves slightly 
rightward at $ \geq 300 M_{\odot}$, 
while the metal--rich sequences curve strongly at above several tens of 
$M_{\odot}$.
This is because the core--halo structure develops well in metal--rich stars, 
and the effective temperature is lower than that of the metal--poor 
star with the same mass. 
Therefore, the effective temperature depends strongly on the 
metal contents, while the luminosity depends 
weakly on the metallicity, as shown in figure 4.  

\begin{figure}
\begin{center}
\epsfile{file=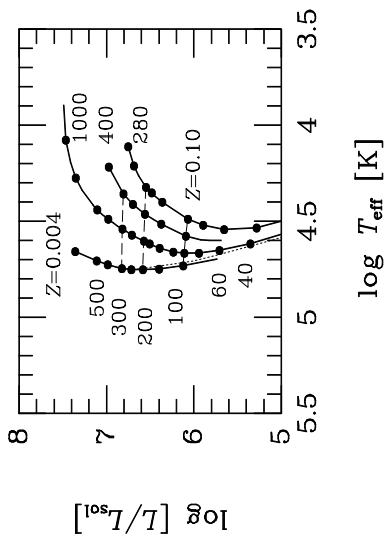,width=5cm,height=7cm }
\end{center}
\caption{
Track of zero--age main--sequence with various metal contents in the H--R
diagram. The solid curves denotes the main--sequence with $X=0.7$, and
$Z=0.004,0.02,0.05$, and 0.10, from left to right. The stellar mass is noted
next to the dot in units of  $M_\odot$.
The dotted curve  up to $240 M_\odot$
denotes ZAMS ($Z=0.03$) with old opacity  (Maeder 1980).
}
\end{figure}

Figure 5 depicts $U$--$V$ curve of the stars of 40, 200, and 
as the extreme massive case  1000 
$M_{\odot}$, to demonstrate 
the difference of in the interior structures of the Type--I and Type--II 
solutions, where $U$ and $V$ are the homologous invariants, defined by

\begin{equation}
\label{eq:U}
$$U=4 \pi r^3 \rho /M_r$$ 
\end{equation}

\noindent and 
\begin{equation}
\label{eq:V}
$$V=GM_r \rho /(rP)$$.
\end{equation}

\noindent
Here, $U$ represents  the density divided by the mean density, and $V$ is 
the ratio of the gravitational energy to the thermal energy.  
A Type--I solution, such as of the $40 M_{\odot}$ main--sequence star, 
shows almost a monotonic increase of $V$ from the center to the photosphere,  
whereas the Type--II solutions of 200 and 1000 $M_{\odot}$ make a deep dip 
or a loop around $V \sim 2$. The decrease of  
$V$ toward the dip is caused by a quick decrease in the density 
outward, which corresponds to a steep rise in the opacity. 
After the loop, the opacity begins to decrease, which keeps the 
density at a relatively large value, and then $V$ increases again. 
Note that such a loop structure is a characteristic property 
of red giants with hydrogen--shell burning, which has an extended 
envelope around a dense core (Hayashi et al. 1962). 

\begin{figure}
\begin{center}
\epsfile{file=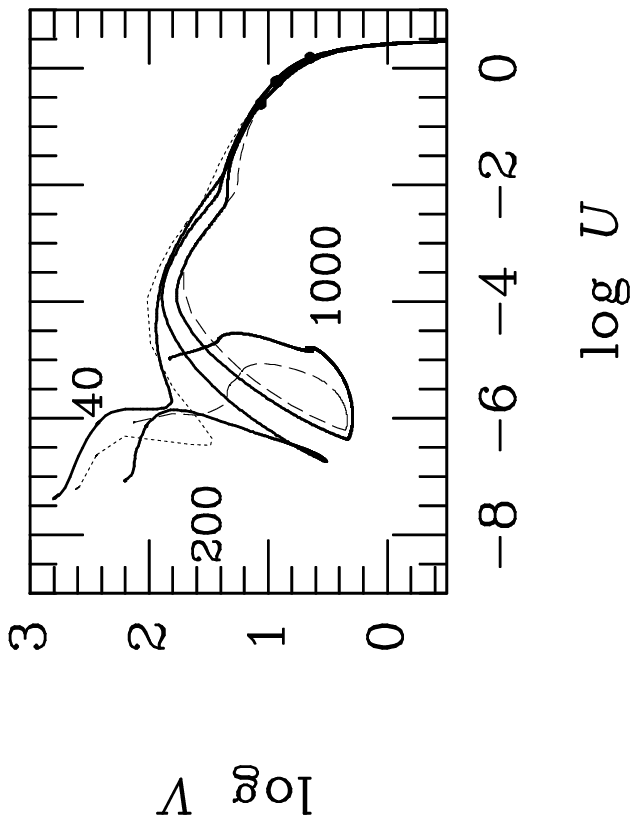,width=7cm,height=5cm }
\end{center}
\caption
{
U--V curves for main--sequence stars of 40, 200,
and 1000 $M_{\odot}$ with $X=0.7$ and $Z=0.02$, where $U=  
4 \pi r^3 \rho /M_r$ and $V=GM_r \rho /(rP)$. The stellar mass is   
attached beside the curve. Two models of 200 $M_{\odot}$
with $Z=0.004$ (dotted curve) and $Z=0.10$ (dashed curve) are also shown.
The outer edge of the convective core is indicated by filled circles
for 1000, 200, and 40 $M_{\odot}$ from left to right.
}

\end{figure}

\section
{Discussion}


Figure 6 shows ZAMS tracks for various chemical composition. 
The thick and thin curves denote the ZAMS with $X=0.7$ and $Z=0.0$, 
respectively. One additional sequence of $X=0.35$ and $Z=0.05$ is 
also shown for a comparison.  Table 1 shows that a hydrogen--deficient star 
($X=0.35, Z=0.05$) has a higher luminosity and a lower temperature, 
compared with that of a hydrogen--rich star ($X=0.7, Z=0.05$) 
of the same mass. This makes a good contrast to the weak dependence 
of the luminosity on $Z$; i.e., 
the stellar luminosity is almost independent of $Z$ 
for a given mass. The difference between helium--rich stars ($X=0.35$) 
and solar composition stars ($X=0.7$) in luminosity and 
temperature  is  $\Delta \log L= 0.22$ and $ \Delta T_{\rm eff} 
=-0.26$, for 150 $M_{\odot}$, 
and $\Delta \log L= 0.195$, and $ \Delta T_{\rm eff}=-0.50$
for $200 M_{\odot}$. In other words, observational information on helium 
enrichment is important when we derive the stellar masses from 
the observed luminosities. 

We now compare our theoretical ZAMS tracks in the H--R diagram with 
several very luminous stars which were recently discovered. 
Figure 6 also depicts the position of several very bright stars 
[$\log L/(L_{\odot}) > 6$] in our Galaxy, the LMC, and the SMC.  
The distribution of the LMC and the SMC stars (filled symbols) 
is consistent with our ZAMS curve of $Z=0.04$, because Magellanic stars 
are known to be metal poor. 
%
%
A very massive LMC star, Melnick 42, denoted by the filled circle, 
has been examined in detail by Pauldrach et al. (1994), who derived the 
stellar parameter to be $\log(L/L_{\odot})=6.6$, 
$T_{\rm eff}=50500$ K, $Z=Z_{\odot}/4$, 
and $M/M_{\odot}=150$, from model fitting of non--LTE UV spectrum with 
HST UV spectrum. The position of this star in figure 6 (filled circle)
is close to the ZAMS of $Z=0.004$ and quite consistent with our 
aspect. From our sequence of 
$X=0.7$ and $Z=0.004$, its mass is estimated to be $\sim 200 M_{\odot}$;  
if the star is helium--rich, the mass is slightly smaller than this
value, as mentioned above. 

The distribution of these Magellanic Cloud stars seems to be on the whole
slightly leftside to that of the galactic stars (open square). This is 
consistent with the metal deficiency of the Magellanic Clouds stars,  
although the data number is not sufficient for making any definite
statement, and some of which possibly have 
left the main--sequence to a redward evolutionary excursion. 

This figure also shows four massive stars in the galactic center 
(open triangles). 
Najarro et al. (1997) have analyzed He {\footnotesize I} lines of 
these stars by a non--LTE radiative--transfer method for a pure H--He 
spherical atmosphere. They showed that all of them are strongly He 
enriched (He/H $ > 1$). 
Because these IRS stars are in the galactic center, they are possibly 
enriched in metal as well. In the H--R diagram, they are located 
close to our theoretical curve of $X=0.35$ and $Z=0.05$ in the 
upper--half part that represents core--halo structure. 
Therefore, if their interior structure 
is not far from that of our uniform model, 
we can expect that their low effective temperature may be attributed to  
the core--halo structure. 

\begin{figure}
\begin{center}
\epsfile{file=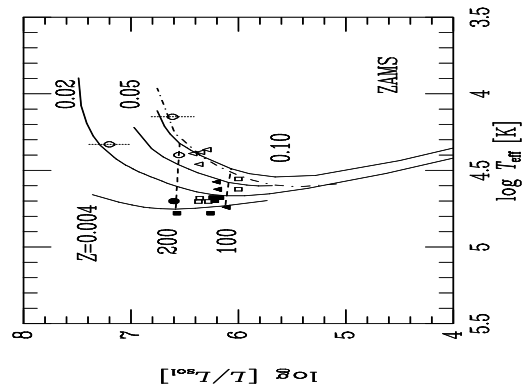,width=7cm,height=5cm }
\end{center}
\caption
{
Theoretical ZAMS ($X=0.7$) plotted in the H--R diagram (thick curves).
The heavy element content of each sequence is assumed to be
$Z=0.004,0.02,0.05$, and 0.10, from left to right.
The two horizontal dashed curves connect the points of 200 $M_{\odot}$
(above) and of 100 $M_{\odot}$ on each ZAMS track.
The uppermost point of each curve corresponds to 800, 1050,
400, and  280 $M_{\odot}$.
One additional track of helium--rich ZAMS ($X=0.35, Z=0.05$)
for stars from 20 $M_{\odot}$ to 200 $M_{\odot}$
is denoted by the dot-dashed curve.
Very luminous stars in the Galaxy (open symbols) and in the
Magellanic Clouds (filled symbols) are also plotted.
The two open circles with the error bar denote the two possible
positions of the Pistol Star
(Figer et al. 1998), the open circle $\eta$ Car (Humphreys,
Davidson 1994), and
the filled circle Melnick 42 (Pauldrach et al. 1994).
Several O stars in the Galaxy, the SMC, and the LMC (Puls et al.
1996) are plotted by the open squares, the filled triangles,
and the filled squares, respectively.
The four open triangles denote bright He--rich stars
in the galactic center region (Najarro et al. 1997).
}
\end{figure}

Another candidate for the core--halo structure is the Pistol star. 
The luminosity and temperature of the Pistol star has been derived 
by Figer et al. (1998) from a near--infrared observation 
to be two possibilities:
(case L): $\log L(L_{\odot})=6.6 \pm 0.2$ and $\log T_{\rm eff}(K)=
4.15 \pm 0.01$ K and (case H): $\log L(L_{\odot})=7.2 
\pm 0.2$ and $\log T_{\rm eff}({\rm K})= 4.33 \pm 0.01$ K. They have 
analyzed the star to be slightly helium
rich, $n_{\rm He}/n_{\rm H}=0.$13--0.12; This gives $X=0.6$3--0.66 for 
$Z=0.02$--0.05. The position of case L is very close to the curve of 
ZAMS with $X=0.7$ and $Z=0.1$. 
This metallicity seems to be too high, but we may not exclude $Z=0.1$ for
the Pistol star, because it is in the star--forming region in the 
galactic center. 
If the Pistol star is extremely metal--rich, the star can be interpreted 
to be a near zero--age main--sequence with a core--halo structure. 
In the case of $Z=0.0$2--0.05, the star is 
interpolated to be a young star that is 
evolved redward apart from ZAMS.  In any case, the stellar mass is 
estimated to be $\sim 250 M_{\odot}$, which is in a good agreement with 
200--250 $M_{\odot}$ obtained by Figer et al. In case H, 
the star is close to the main--sequence with $Z=0.02$ and the mass 
is estimated to be extremely high, as much as 700 $M_{\odot}$. 

In this way, the low effective temperature of a very luminous star 
can be interpolated in part by the core--halo structure. 
It is difficult, however, to confirm our prediction, because  
we need to distinguish a very young star having a core--halo 
structure from an evolved star. 
Observational estimates of the helium and heavy element contents 
of such stars are useful information as well as a detailed 
evolutional calculation.

Figer et al. (1998) have shown ZAMS tracks up to $300 M_{\odot}$ with 
$Z=0.02$ and 0.04, of which the upper half curves rightward, 
which is qualitatively in good agreement with 
ours. Their temperatures are systematically higher than those of ours 
($\Delta \log T_{\rm eff} \sim 0.1$ at $200 M_{\odot}$), whereas 
the luminosity is the same.  
This difference may be caused by a different treatment of the radiative 
envelope, because they combined an interior solution to a 
steady wind solution with a given density profile, 
while we solved the structure of the entire radiative envelope. 

\begin{figure}
\begin{center}
\epsfile{file=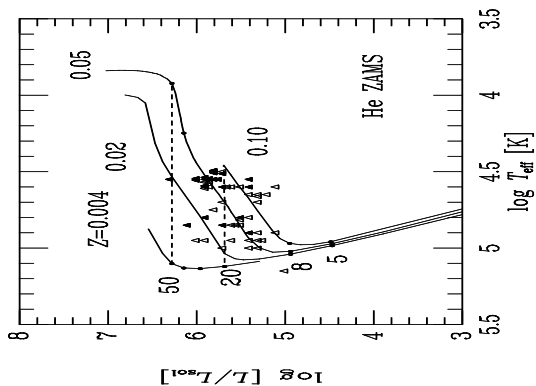,width=7cm,height=5cm }
\end{center}
\caption
{
Same as those in figure 6, but for helium ZAMS ($X=0$).
The stellar mass at the small dots on the curves are, from
lower to upper, 5, 8, 20, 30, 40, and 50 $M_\odot$
for the $Z=0.004$ sequence, and 5, 8, 40, and 50
$M_\odot$ for $Z=0.05$. The mass of the uppermost point   
of each curve is 80, 130, 220, and 20 $M_\odot$ from left to right.
The two horizontal dashed curves connect the points of 20
$M_{\odot}$ (lower) and of 50 $M_{\odot}$ on each track.
The triangles denote the galactic WN stars;
the open triangles represent stars in which hydrogen is not detectable
in their atmosphere, and the filled triangles with hydrogen
(Hamann, Koesterke 1998).
}
\end{figure}

Figure 7 shows theoretical He ZAMS ($X=0$) with different metal
content. The characteristic properties are the same as those of
ZAMS in figure 6, but the He ZAMSs bend at lower stellar masses.
Compared with the helium ZAMS obtained by Langer (1989) with the  
Los Alamos opacity, our sequences are close to his curve
in the lower part ($ < 5 M_{\odot}$), but bend rightward at 
less--massive stars than does Langer's curve. This is naturally 
explained by the difference in the opacity; the large peak of the OPAL 
opacity causes a core--halo structure in less massive stars, which 
causes bending of the main--sequence.

This figure also depicts the position of galactic Wolf--Rayet stars
obtained from a non--LTE radiative transfer calculation for a helium and
nitrogen atmosphere (Hamann, Koesterke 1998). Most of the stars
distribute in the rightside of the curve of $Z=0.02$.
The open triangles denote a star in which hydrogen is not detectable in
the atmosphere. If all of these stars have the same metallicity
as that of the sun, $Z\sim 0.02$, their rightward deviation from the curve
of $Z=0.02$ can be interpreted based on evolutional effects. Langer (1989) has
followed the evolution of helium stars by decreasing the helium content
of the convective core, and showed that massive stars ($M \geq 15
M_{\odot}$) leave the main--sequence redward in the H--R diagram, 
while less massive stars ($M < 10 M_{\odot}$) move leftward. Considering 
the difference in the opacity, i.e., the large peak in the OPAL opacity 
is not present in the Los Alamos opacity, we can naturally expect that 
stars of mass $> 10 M_{\odot}$ move rightward in the H--R diagram.

The distribution of filled triangles, that denotes stars with
hydrogen 
in the atmosphere,  seems to be weighted to the lower temperature side, 
which is explained by the difference in the opacity. When a helium  
core--burning star has a hydrogen--rich atmosphere, the star intends to
have a more extended radiative envelope due to an increase of the
opacity in its surface region.

A few stars locate to the leftside of the curve of $Z=0.02$, (e.g., a
star at $\log T_{\rm eff} = 4.95$ and $\log L/L_{\odot} = 6$). 
This may be attributed in part to the difference in the definition 
of the temperature between ours and Hamann and Koesterke, in which their 
temperature, depicted here, is defined as that of the bottom   
of the expanding envelope, where the optical depth is 20. Therefore,
the temperature that should be compared with our curve may be slightly 
smaller than these values. In the presence of a wind mass--loss, however,
a rigid definition of the surface temperature will be very complicated and 
difficult to compare with ours; such detailed examinations are 
beyond the scope of the present paper.
Within such ambiguities, we therefore conclude that the bending of He
ZAMS is consistent with the observed data of WN stars.

\section{Conclusions}

1. We present the internal structure of chemically homogeneous stars 
with central nuclear burning. Massive stars develop a core--halo structure 
($>100 M_{\odot}$ for MS star, and $>15 M_{\odot}$ for helium MS star) 
with a very extended radiative envelope, 
whereas less massive stars do not. This structure change is caused by 
the large peak of the OPAL opacity, which is  
more prominent in more massive metal--rich stars.

2. The upper part of the main--sequence curves redward in the H--R diagram, 
because the core--halo structure develops well 
in more massive stars and the surface temperature decreases along with 
an increase in the stellar mass. 
The critical mass of main--sequence turning to 
rightward is $200 M_{\odot}$ for 
$Z=0.004$, $100 M_{\odot}$ for $Z=0.02$, and $70 M_{\odot}$ for
$Z=0.05$ for ZAMS, and $30 M_{\odot}$ for
$Z=0.004$, and $15 M_{\odot}$ for $Z=0.02$ for He ZAMS. Because the 
Population I main--sequence more strongly bends than does Population II, 
very young massive main--sequence stars are expected to distribute in the 
order of the metal content in the upper H--R diagram. 

3. The distribution of observed very massive stars in the H--R diagram is 
consistent with our theoretical main--sequences. 
A ZAMS star of extreme Population I has a very low $T_{\rm eff}$ despite  
its young age, which possibly leads to a misclassification as an evolved 
star. The large radius, owing to the core--halo structure of very young 
MS star, must be distinguished from the redward excursion from  
evolutionary effects by observational information on the chemical composition.

\par
\vspace{1pc}\par
We appreciate a referee, N. Langer, for valuable comments. 
MI and MK are grateful for the hospitality of the College of Arts and 
Sciences, the University of Tokyo. 
This research has been supported in part by a Grant--in--Aid for
Scientific Research (09640325)
of the Japanese Ministry of Education, Science, Sports, and Culture. 

\section*{References}
\small

\re
Figer D.F., Morris M., McLean I.S., 1996, in The Galactic Center, 4th ESO,
CTIO Workshop, ed R.\ Gredel, ASP Conf.\ Ser.\ 102,  263

\re
Figer D.F., Najarro F., Morris M., McLean I.S., Geballe T.R., 
Ghez A.M., Langer N.\ 1998, ApJ.\ 506, 384

\re
Glatzel W., Kiriakidis M.\ 1993, MNRAS 262, 85


\re
Glatzel W., Kiriakidis M.\ 1998, MNRAS 295, 251

\re
Hamann W.-R., Koesterke L.\ 1998, A\&A 333, 251

\re 
Hayashi C., Hoshi R., Sugimoto D.\ 1962,  Prog.\ Theor.\ Phys.\ 22, 1 
\re
Heger A., Langer N.\ 1996, A\&A 315, 421

\re 
Humphreys R. M., Davidson K.\ 1994, PASP 106, 1025  

\re
Iglesias C.A.,  Rogers F. J., 1996, ApJ\ 464, 943


\re
Kato M.\ 1980, Prog.\ Theor.\ Phys.\ 64, 847
\re
Kato M.\ 1985, PASJ 37, 311

\re
Kato M.\ 1986, Ap\&SS 119, 57

\re
Kato M., Iben I.Jr 1992, ApJ\ 394, 305
\re

\re
Langer N.\ 1989, A\&A 210, 93 

\re
Massey P., Hunter D.A.\ 1998, ApJ\ 493, 180

\re
Maeder A.\ 1980, A\&A 92, 101
\re
Nagata T., Hyland A.R., Straw S.M., Sato S., Kawara K.\ 1993, ApJ\ 406, 501

\re
Nagata T., Woodward C.E., Shure M., Kobayashi N.\ 1995, AJ\ 109, 1676

\re
Najarro F., Krabbe A., Genzel R., Lutz D., Kudritzki R.P., 
Hillier D. J.\ 1997, A\&A 325, 700

\re
Pauldrach A.W.A., Kudritzki R.P., Puls J., Butler K., Hunsinger J.\
  1994, A\&Ap 283, 525
  
\re
Puls J., Kudritzki R.-P., Herrero A., Pauldrach A.W.A., Haser
S.M., Lennon D.J., Gabler R., Voels S.A.\ et al.\ 
\ 1996, A\&A 305, 171

\re
Schaller G., Schaerer D., Meynet G., Maeder A.\ 1992, A\&AS 96, 269



\re
Stothers R.B. \& Chin C-w.\ 1996, ApJ 468, 842
\re
Stothers R.B. \& Chin C-w.\ 1997, ApJ 489, 319    

\end{document}